\documentclass[prc,preprint,showpacs,amsmath,amssymb]{revtex4}
\usepackage{graphicx}
\usepackage{dcolumn}
\usepackage{bm}
\usepackage{color}
\begin{document}

\title{Identifying resonances with wave-packet dynamics}

\author{Alexis Diaz-Torres and Jeffrey A. Tostevin}

\affiliation{Department of Physics, University of Surrey, Guildford GU2 7XH, UK}

\date{\today}

\begin{abstract}
A new method for the study of resonant behavior - using wave-packet dynamics - is
presented, based on the powerful window operator technique. The method is illustrated
and quantified by application to the astrophysically-important example of low-energy
$^{12}$C + $^{12}$C collisions. For this selected, potential model test case, the
technique is shown to provide both resonance energies and widths in agreement with
alternative methods, such as complex-energy scattering-matrix pole searches and
scattering phase-shift analyses. The method has a more general capability to study
resonance phenomena across disciplines, that involve particles temporarily trapped
by potential pockets.
\end{abstract}


\maketitle

\section{Introduction}

Resonance phenomena, where particles are temporarily trapped by attractive potential
pockets, are ubiquitous in many fields of physics, chemistry, molecular biology and
technology. For instance, shape-type or potential resonances appear in resonance
tunneling in quantum wells and quantum wires \cite{Bhatta} as well as in the rapid
neutron-capture process that is crucial for heavy element creation in the Universe
\cite{Rolfs}.

To describe such resonance phenomena quantitatively, various methods have been developed
using Hermitian and non-Hermitian quantum mechanics. These methods include the location
of poles of the scattering matrix (S-matrix) in the complex energy plane, analysis of the
density of states spectra, the asymptotes of continuum eigenfunctions, the scattering
phase shifts, complex scaling methods, and filter diagonalization methods. These methods
are comprehensively reviewed in Ref. \cite{Nimrod}. Resonances can also be determined by
analytic continuation of eigenvalues from the stabilization graph into the complex energy
plane \cite{Nimrod2} and by using wave-packet dynamics \cite{Nimrod3}.

The present paper provides a new method for describing potential resonances, using
wave-packet dynamics \cite{Tannor,Boselli, ADTWiescher} along with the powerful window
operator method \cite{Schafer}. As a topical test case, and so as to allow direct,
quantitative comparisons with alternative techniques, we use a simplified potential
model dynamical description of $^{12}$C + $^{12}$C nuclear collisions at bombarding
energies near to the Coulomb barrier. The results of the present method can be compared
directly with those obtained using complex energy plane S-matrix pole searches and
analyses of the scattering phase shifts \cite{Merzbacher}. We note that many of the
alternative methods, mentioned above, will actually be more efficient in determining
such potential resonances in the test case of the present paper. However, the one-dimensional
potential model of $^{12}$C + $^{12}$C collisions allows illustration of the key ideas
of the proposed method that describes resonances dynamically. More generally, in the
physical description of nuclear collisions and reactions, for instance the fission and
fusion of heavy ions, an understanding of the dynamical formation and decay of intermediate
(nuclear molecular) structures is crucial \cite{ADTWiescher,Tumino,Goldstone,Tudora}.
Usually, multi-dimensional potential pockets support these short-lived structures, whose
description in terms of stationary-state methods is challenging as the boundary conditions
are unclear. In such contexts, the present method is useful. Here, we present both a
description of the model problem and a critical evaluation of the proposed method
and its results, followed by a summary.

\section{Model and Methods}

\subsection{General aspects of the time-dependent wave-packet method}

The time-dependent wave-packet method involves three steps \cite{ADTWiescher,Tannor,Boselli}:
\begin{itemize}
\item[(i)] the definition of the initial wave function $\Psi(t=0)$,
\item[(ii)] the propagation $\Psi(0) \to \Psi(t)$, dictated by the time evolution operator,
$\exp (-i \hat{H} t/\hbar )$, where $\hat{H}$ is the total Hamiltonian that is time-independent,
\item[(iii)] the calculation of observables (e.g., spectra) from the time-dependent wave function, $\Psi(t)$.
\end{itemize}

The wave function and the Hamiltonian are represented in a grid. In the present work,
for simplicity, these are considered a function of only the internuclear distance, $R$.

\subsection{The Hamiltonian}

\begin{figure}
\centering
\includegraphics[width=8.5cm,clip]{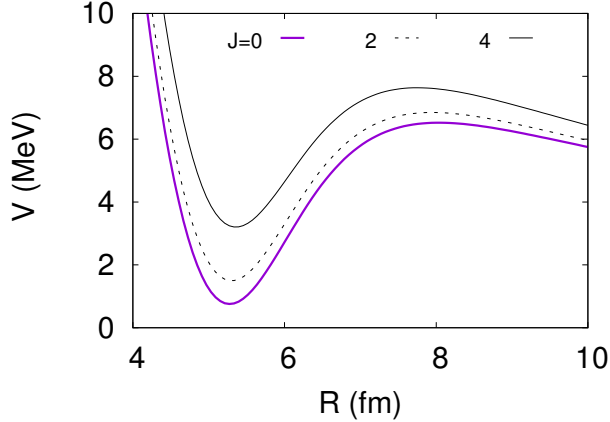}
\caption{(Color online) Real $^{12}$C + $^{12}$C total potentials for the $J=0, 2, 4$ partial
waves \cite{suppl}.} \label{fig-1}       
\end{figure}

The Hamiltonian, $\hat{H} = \hat{T} + \hat{V}$, is formed by the radial kinetic energy operator,
$\hat{T} = -\frac{\hbar^2}{2 \mu}\frac{\partial^2}{\partial R^2}$, and a real total potential
energy, $\hat{V} = \frac{\hbar^2 J(J+1)}{2 \mu R^2} + U(R)$, which includes a centrifugal component
with a total angular momentum $J$. Here $\mu$ denotes the reduced mass and $U(R)$ is comprised
of the nuclear and Coulomb interactions. Fig.~\ref{fig-1} shows the real total potentials for
$^{12}$C + $^{12}$C for the $J=0, 2, 4$ partial waves, as calculated in Ref. \cite{ADTWiescher}.
These potentials were determined by folding the collective potential-energy landscape (including
the centrifugal energy) of non-axial symmetric di-nuclear configurations with the probability
density of the ground-state wave-function of the two deformed, colliding $^{12}$C nuclei \cite{suppl}. The
heights of the Coulomb barriers are $6.5$~MeV ($J=0$), $6.8$~MeV ($J=2$), and $7.6$~MeV ($J=4$)
and their radii are $8$~fm, $7.9$~fm and $7.8$~fm, respectively. Resonance states are formed
in the attractive pockets of these potentials, as is shown below.

\subsection{Initial wave function and time propagation}

The initial wave function, $\Psi_{0}(R)$, is described by a Gaussian wave-packet, centered
at a large internuclear distance $R_0$ and having a spatial width $\sigma$. This initial
wave-packet is boosted toward the potential barriers, shown in Fig. \ref{fig-1}, with an
average wave-number $K_0$ appropriate to the kinetic energy for the mean total incident
energy $E_0$. That is,
\begin{equation}
\Psi_0(R) = \frac{1}{\pi^{1/4}\sqrt{\sigma}}\,exp \Big[ -\frac{(R-R_0)^2}{2\sigma^2}\Big]
\,e^{-iK_0 (R-R_0)}.
\label{eq1}
\end{equation}

In the following calculations, an initial wave-packet with $R_0 = 400 \, \textnormal{fm}$,
$\sigma = 10$ fm and the mean total incident energy around the heights of the Coulomb barriers,
$E_0 = 6 \, \textnormal{MeV}$, is assumed. These parameters do not affect the resonance energies
calculated \cite{Boselli}, provided (i) the initial wave-packet is fully contained in the
position and wave-number grids, (ii) its amplitude at the grid edges is negligible, and
(iii) the initial position $R_0$ is larger than the classical turning points for wave-numbers
in the interval $(-K_0 \pm 3/\sigma)$. The wave-packet is represented on a Fourier radial grid
\cite{Kosloff2}: $R = 0-1000 \, \textnormal{fm}$ with $2048$ evenly spaced points.

The time propagation of the wave function was carried out using the Chebyshev propagator
\cite{Kosloff2} for the evolution operator, as is described in Appendix C of Ref.
\cite{ADTWiescher}. In this work the time-step used was $\Delta t = 10^{-22}$ s, for which
the normalization of the wave function is preserved with an accuracy of $\sim 10^{-14}$.
As was stated above, for the sake of simplicity and for ease of comparisons with alternative
methods, other dynamical processes, such inelastic scattering of the $^{12}$C nuclei, are
neglected. The initial and time-evolved wave-packets contain a range of translational
energies, so the calculation of observables requires an energy projection method.

\subsection{The energy projection method}

Within the window operator approach, the energy spectrum of a time-dependent numerical wave
function, $|\Psi \rangle$, is $\mathcal{P}(E_k) \,=\, \langle \Psi | \hat{\Delta} |\Psi \rangle $,
where $\hat{\Delta}$ is the window operator with centroid energy $E_k$ \cite{Schafer}:
\begin{equation}
\hat{\Delta}(E_k,n,\epsilon)\, \equiv \, \frac{\epsilon^{2^n}} {(\hat{H}\, - \, E_k)^{2^n}\, +
\, \epsilon^{2^n}} .
\label{eq5}
\end{equation}
Here, $\hat{H}$ is the system Hamiltonian, and $n$ determines the shape of the window function.
$\mathcal{P}(E_k)$ represents the probability of finding the system, in state $|\Psi \rangle$,
in the energy window $E_k\pm \epsilon$. As $n$ is increased, the shape of the window function
rapidly becomes rectangular with very little overlap between adjacent energy bins with centroid
energies $E_k$, the bin width remaining constant at $2\epsilon$ \cite{Schafer}. The spectrum
is constructed for a set of $E_k$ where $E_{k+1}=E_k + 2\epsilon$. In this work, $n=2$ and
$\epsilon$ will be specified in each of the calculations below. Solving two successive linear
equations for the vectors $|\chi_k \rangle$:
\begin{equation}
(\hat{H}\, - \, E_k \, + \, \sqrt{i}\,\epsilon)(\hat{H}\, - \, E_k \, - \, \sqrt{i}\,\epsilon)
\, |\chi_k \rangle \,=\, \epsilon^2 \, |\Psi \rangle, \label{eq6}
\end{equation}
yields $\mathcal{P}(E_k) \,=\, \langle \chi_k |\chi_k \rangle $. Thus, the vectors $|\chi_k
\rangle$ represent bin states over the energy range $E_k\pm \epsilon$.

\subsection{Potential resonances from wave-packet dynamics}

\begin{figure}
\centering
\includegraphics[width=8.5cm,clip]{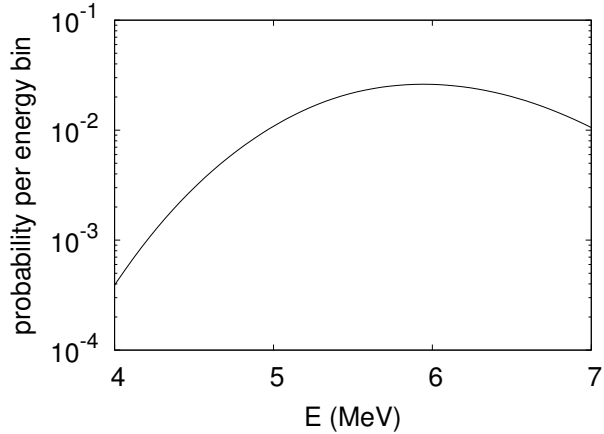}
\caption{The energy spectrum of the initial wave-packet with $E_0 = 6$ MeV. The bin width is
$2\epsilon=50$~keV and the maximum radius of the grid is $R_{max} = 1000$~fm.}
\label{fig-2}       
\end{figure}

\begin{figure}
\centering
\includegraphics[width=8.5cm,clip]{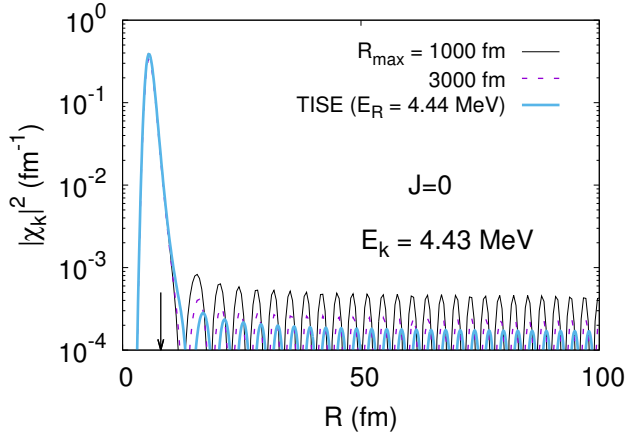}
\caption{(Color online) The radial probability density of a normalized resonant bin state from
Eq. (\ref{eq6}), for $E_k = 4.43$~MeV and $\epsilon=1$~keV, using grids with $R_{max} = 1000$
and $3000 $~fm. The part, $\tilde{|\Psi\rangle}$, of the time-dependent wave-packet, $|\Psi
\rangle$, that is inside the radius of the $J=0$ Coulomb barrier (arrow) in Fig. \ref{fig-1},
is used in the source term of Eq. (\ref{eq6}). The position of the body of the recoiling $|\Psi
\rangle$, with $E_0 = 6$ MeV, is at ${\cal R}=25$ fm. The resonant bin state from Eq. (\ref{eq6}),
with extended grids, approaches the resonant, continuum-scattering state obtained by direct
integration of the time-independent Schr\"odinger equation (TISE), indicating that $|\chi_k\rangle$
correctly approximate the continuum scattering states.}
\label{BINS}
\end{figure}

The key idea of the present method for determining resonances from wave-packet dynamics is
to calculate the energy spectrum of that part of $|\Psi \rangle$ that is inside the radius
of the $J$-dependent Coulomb barriers in Fig. \ref{fig-1}. This part, denoted $\tilde{|\Psi
\rangle}$, is first normalized to unity, and Eq. (\ref{eq6}) is then solved using $\tilde{
|\Psi \rangle}$ in the source term. At any given time, the \textit{effective} spectrum is:
\begin{equation}
\mathcal{P}(E_k) \, \equiv\, \frac{\tilde{\langle \Psi |} \hat{\Delta} \tilde{|\Psi \rangle}}
{\langle \Psi_0 | \hat{\Delta} |\Psi_0 \rangle}.
\label{eq7}
\end{equation}
The denominator here, the energy spectrum of the initial wave-packet, displayed in Fig.
\ref{fig-2}, is employed to remove the dependence on the relative weights of incident energies
in the initial wave-packet. The effective spectrum is normalized to unity, revealing structures
due to potential resonances. This deduced spectrum converges very quickly as the body of the
recoiling wave-packet moves away from the potential pocket, and long before that body approaches
the external edge of the radial grid. Hence, the $\tilde{|\Psi\rangle}$ is not contaminated
by waves reflected at the external edge of the grid, else a complex absorbing potential would
have to be used at that edge. However, the bin states $|\chi_k\rangle$ of Eq. (\ref{eq6}) are
subject to the boundary condition due to the finite extent, $R_{max}$, of the numerical grid of
the one-dimensional radial box. The discrete representation of the radial kinetic energy operator
uses Eq. (A8) of Ref. \cite{Miller}. Fig. \ref{BINS} shows the normalized resonant bin state, for
$J=0$ and $E_k= 4.43$~MeV, when the position, ${\cal R}$, of the body of the recoiling wave-packet
is at ${\cal R} = 25$ fm. The radial probability density of this bin state inside the radius of the
Coulomb barrier (arrow) is essentially the same for grids with $R_{max}=1000$ and $3000$ fm, while
the (small) amplitude of the long-range oscillatory behavior reduces significantly as the grid is
extended. This resonant bin state also approaches the scattering state obtained by direct
integration of the time-independent Schr\"odinger equation (TISE), the thick solid line in Fig.
\ref{BINS}, as the grid is extended, indicating that the $|\chi_k\rangle$ approximate the
continuum of scattering states.

\section{Results}

\begin{figure}
\centering
\includegraphics[width=8.5cm,clip]{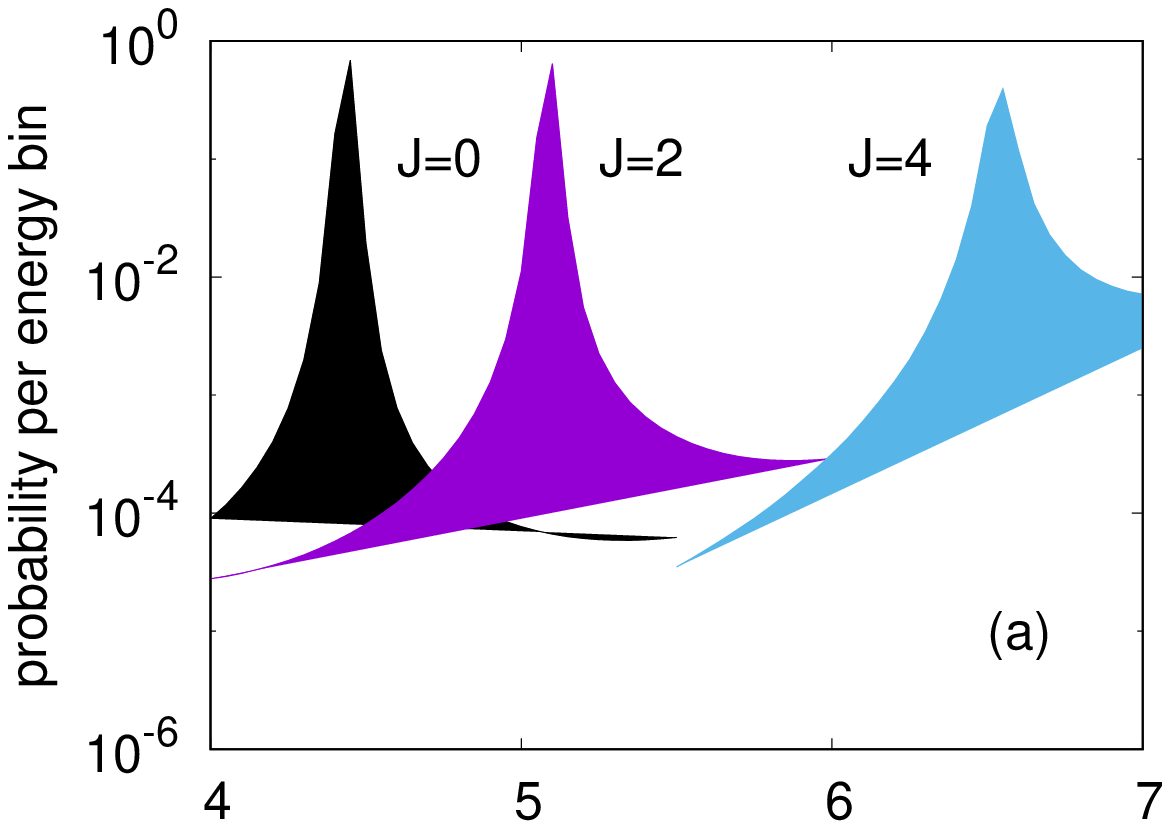} \\
\includegraphics[width=8.5cm,clip]{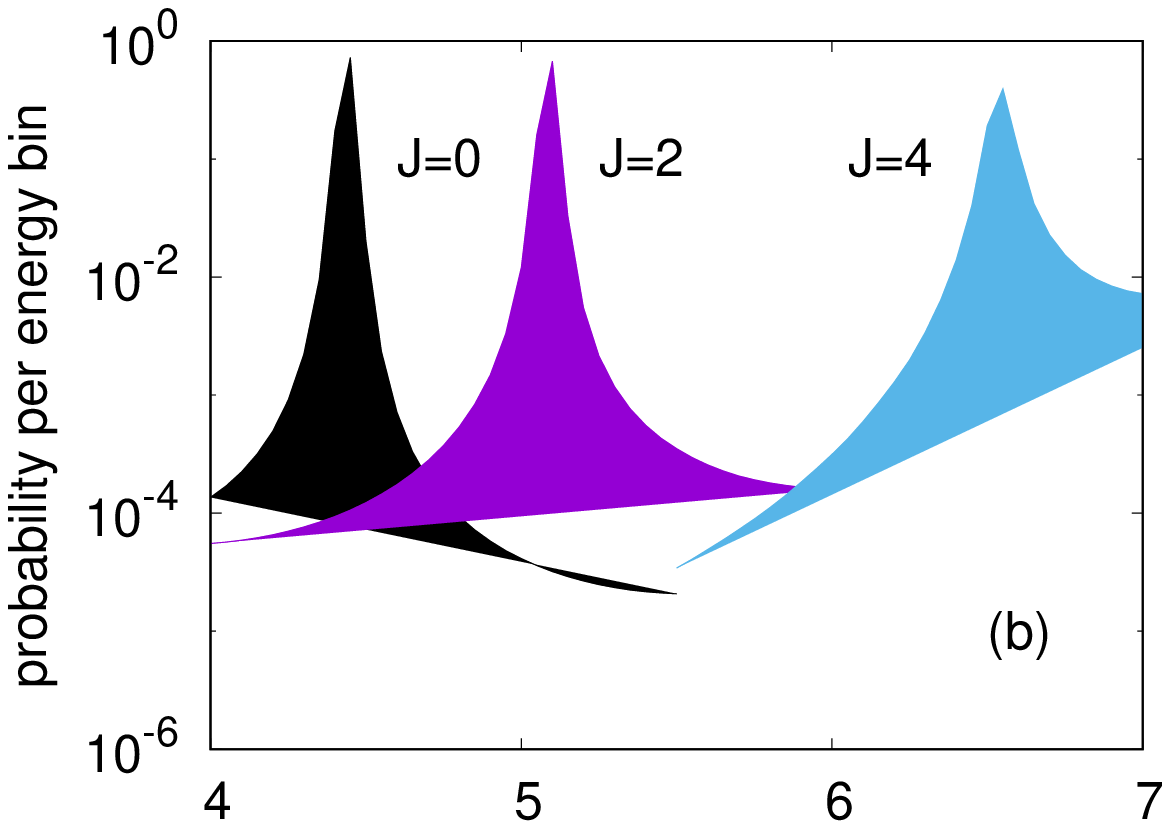} \\
\includegraphics[width=8.5cm,clip]{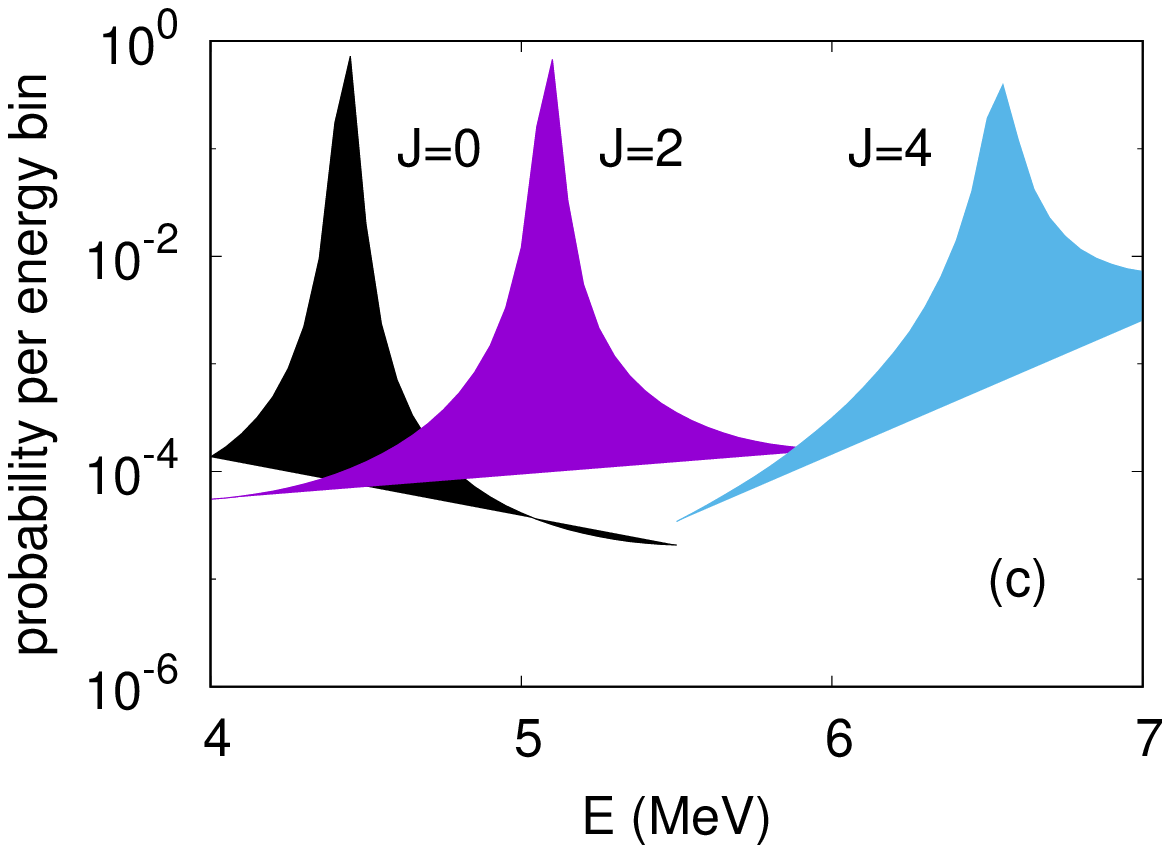}
\caption{(Color online) Effective energy spectra for the part of the time-dependent wave
function inside the radius of the corresponding Coulomb barriers in Fig. \ref{fig-1}. The
bin width is $2\epsilon = 50$~keV and the maximum radius of the grid is $R_{max} = 1000$ fm.
Three instances are shown, when the position ${\cal R}$ of the body of the recoiling wave-packet,
with $E_0 = 6$ MeV, is: (a) near the distance of minimal approach, (b) is at ${\cal R}=25$
fm, and (c) is at ${\cal R}=50$ fm. The strength of these resonances converges rapidly as
the recoiling wave-packet moves away from the Coulomb barriers.} \label{fig-3}
\end{figure}

Figure \ref{fig-3} shows the effective energy spectra of the part of the time-dependent
wave-packet, with $E_0 = 6$ MeV and a given $J$, that is inside the radius of the Coulomb
barriers of Fig. \ref{fig-1}. Three resonant structures are clearly visible in these spectra,
whose strengths converge rapidly as the position of the body of the recoiling
wave-packet moves away from the Coulomb barrier region and the non-resonant energy backgrounds
in this internal region gradually decline.

\begin{figure}
\centering
\includegraphics[width=8.5cm,clip]{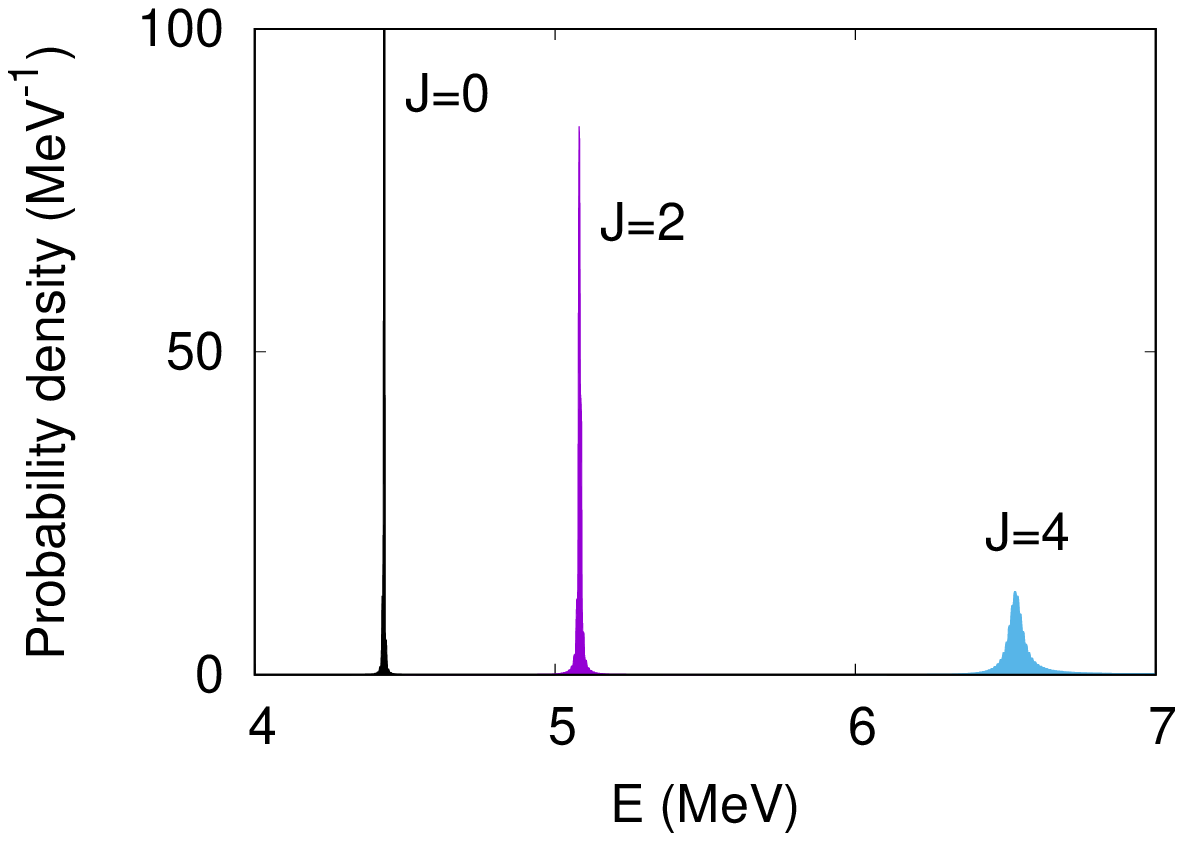}
\caption{(Color online) Probability density function, $\mathcal{P}(E_k)/ 2\epsilon$, for bins
of $2\epsilon=2$~keV and ${\cal R}=25$~fm. The maximum radius of the grid is $R_{max} = 3000$ fm.}
\label{fig-4}       
\end{figure}

\begin{figure}
\centering
\includegraphics[width=8.5cm,clip]{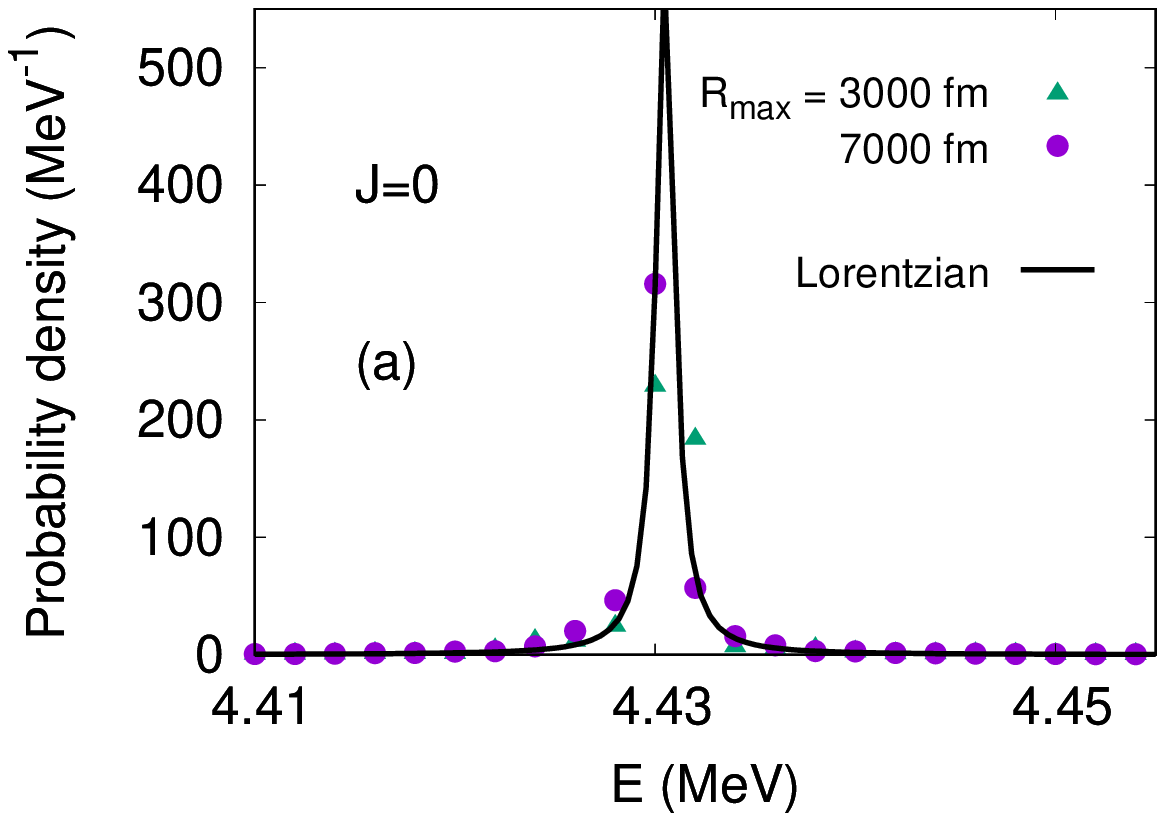} \\
\includegraphics[width=8.5cm,clip]{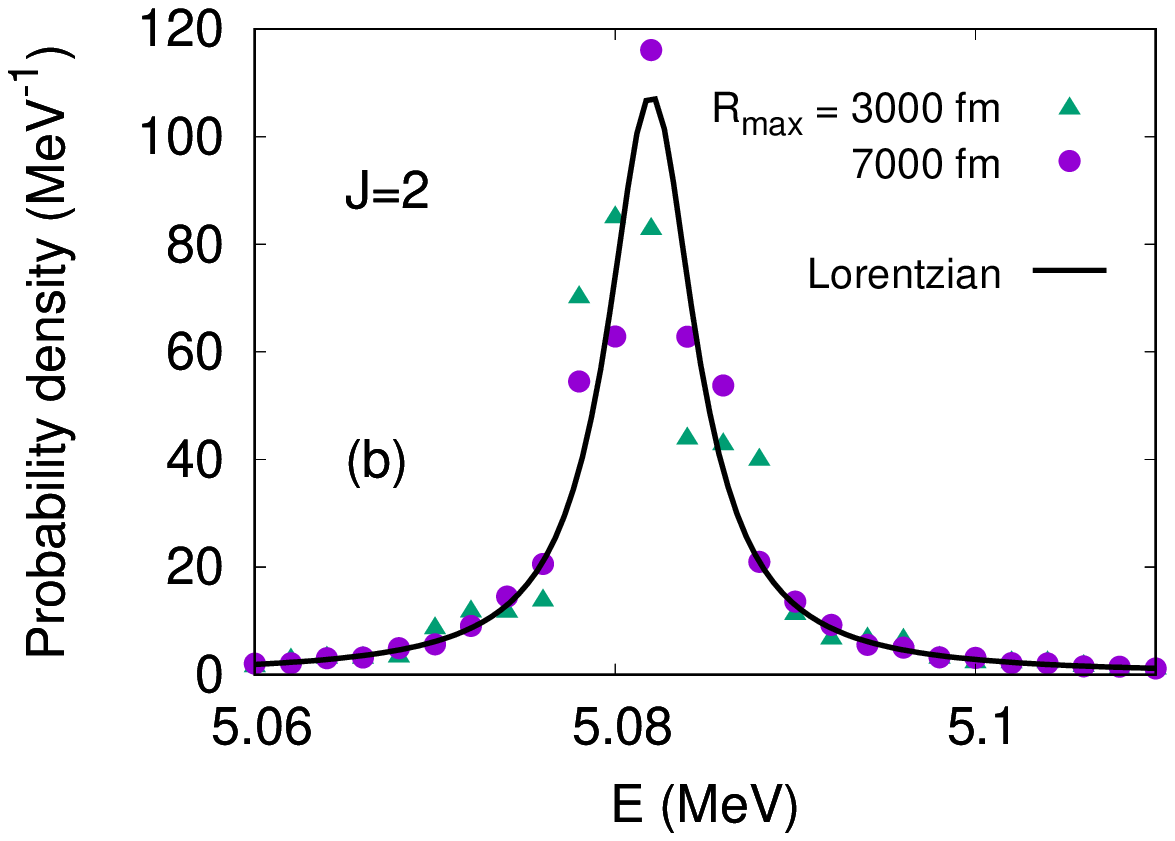} \\
\includegraphics[width=8.5cm,clip]{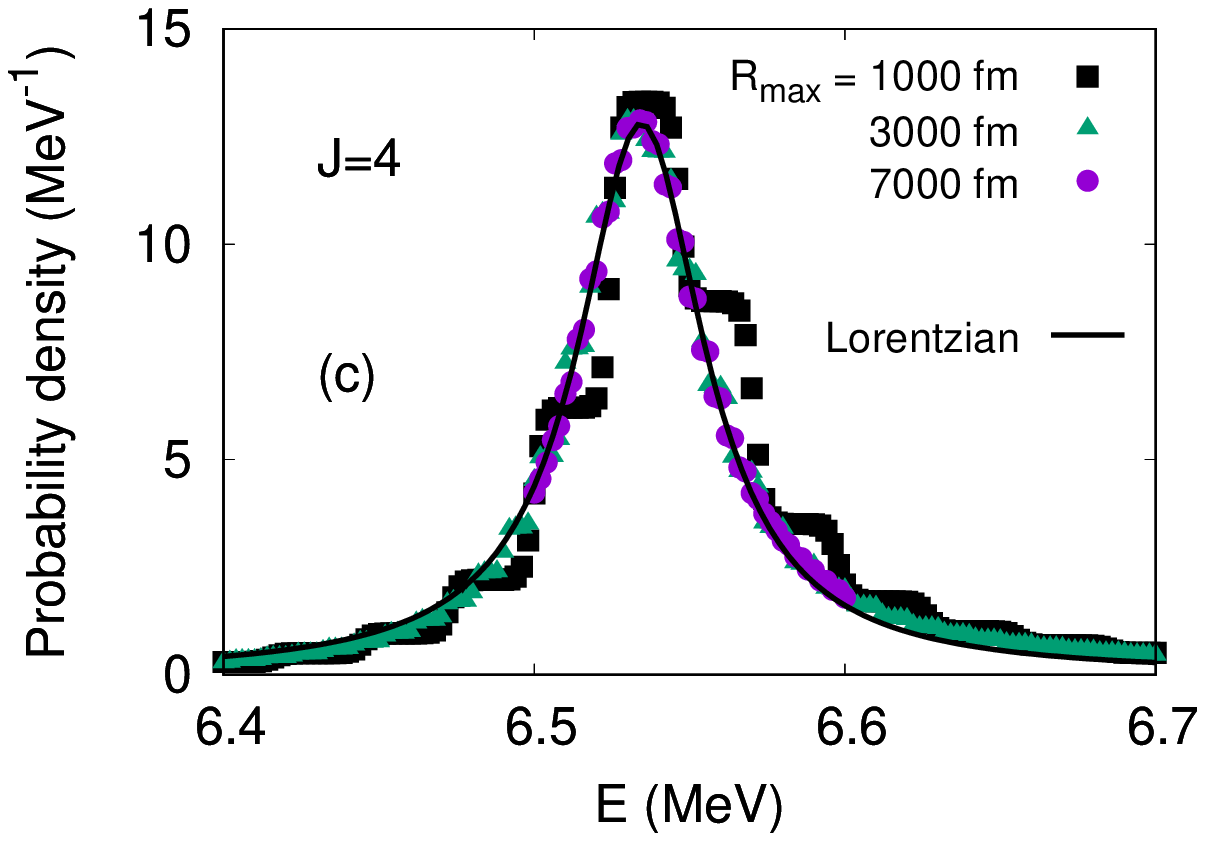}
\caption{(Color online) The same as Fig. (\ref{fig-4}), using various extended grids
(symbols), for (a) $J=0$, (b) $J=2$, and (c) $J=4$. The Lorentzian curves (solid lines)
were fitted to the calculated strength functions for the grid with $R_{max} = 7000$ fm
(solid circles).} \label{fig-5}       
\end{figure}

Figure \ref{fig-4} displays, on a linear scale, a probability density function that results
from dividing a highly resolved effective spectrum by the bin width, i.e., $\mathcal{P}(E_k)/
2\epsilon$. Here the bin widths are $2\epsilon=2$ keV and the wave-packet position is
${\cal R}=25$ fm.

There are two intrinsic energies/resolutions of importance in the numerical calculations.
The first is the chosen bin parameter, $\epsilon$, of the window operator in Eq. (\ref{eq5}).
The second is the spectral resolution of the wave function calculation, the separation
$\Delta E$ between continuum eigenstates of the Hamiltonian imposed by the finite spatial
extent, $R_{max}$, of the numerical grid used. This spectral resolution can be increased,
i.e. $\Delta E$ reduced, to examine narrow resonances, by transferring the numerical wave
functions $|\Psi_{0}\rangle$ and $\tilde{|\Psi \rangle}$ in Eq. (\ref{eq7}), to a larger
grid with the same grid point spacing \cite{Schafer}. The transferred wave functions are
assumed to vanish for radii beyond the original grid. The optimal $R_{max}$ should ideally
be such that the spacing $\Delta E$ between continuum eigenstates is smaller or of the same
order as both the bin width, $2\epsilon$ and the resonance widths of interest. If not, the
computed spectrum reveals artifacts (step-like structures) due to the limited resolution
\cite{Schafer}. A reasonable lower bound on $\Delta E$ in the present case is obtained
assuming free-particle propagation. Since the total potentials involved are less than
0.1 MeV for $R \gtrsim 500$ fm, neighboring eigenstates, for a given $R_{max}$, approximately
satisfy $\Delta K R_{max} = \pi$, with the result that $\Delta E \approx (E/K)(2\pi/R_{max})$,
with $E =\hbar^2K^2/2\mu$. So, for $R_{max}=3000$ fm and $E\approx 4$ MeV, $\Delta E \gtrsim
8$ keV.

This expected improvement in the $\Delta E$ resolution with increasing $R_{max}$ can be observed
in Fig. \ref{fig-5}(c) that presents the probability density for $J=4$ using three different
grid extents (symbols), $R_{max} = 1000, 3000$ and 7000 fm. Figures. \ref{fig-5}(a) and
\ref{fig-5}(b) show this dependence of the probability density on the grid size for the
narrower $J=0$ and $J=2$ resonances, respectively, and the expected improved quality of the
probability density function when increasing the grid size from $R_{max} = 3000$ to 7000 fm.

Fitting the isolated, narrow peaks in Fig. \ref{fig-5} to a normalized Lorentzian:
\begin{equation}
f(E,E_R,\Gamma_R)\, = \, \frac{1}{\pi} \, \frac{\Gamma_R/2} {(E\, - \, E_R)^2\, + \,
(\Gamma_R/2)^2}, \label{eq8}
\end{equation}
both the resonance energies, $E_R$, and their widths, $\Gamma_R$, can be extracted. This
is carried out using the nonlinear least-squares Levenberg-Marquardt algorithm \cite{LM1,LM2}
implemented in the \textit{gnuplot} package. Fig. \ref{fig-5} shows the Lorentzian (solid
line) that fits the probability density for each partial wave, using the grid with $R_{max}
= 7000$ fm (solid circles). For fitting lineshapes of overlapping, broad resonances, the
Pad$\acute{e}$ approximant method \cite{Nimrod2} can be used to determine $E_R$ and $\Gamma_R$.

\begin{figure}
\centering
\includegraphics[width=8.5cm,clip]{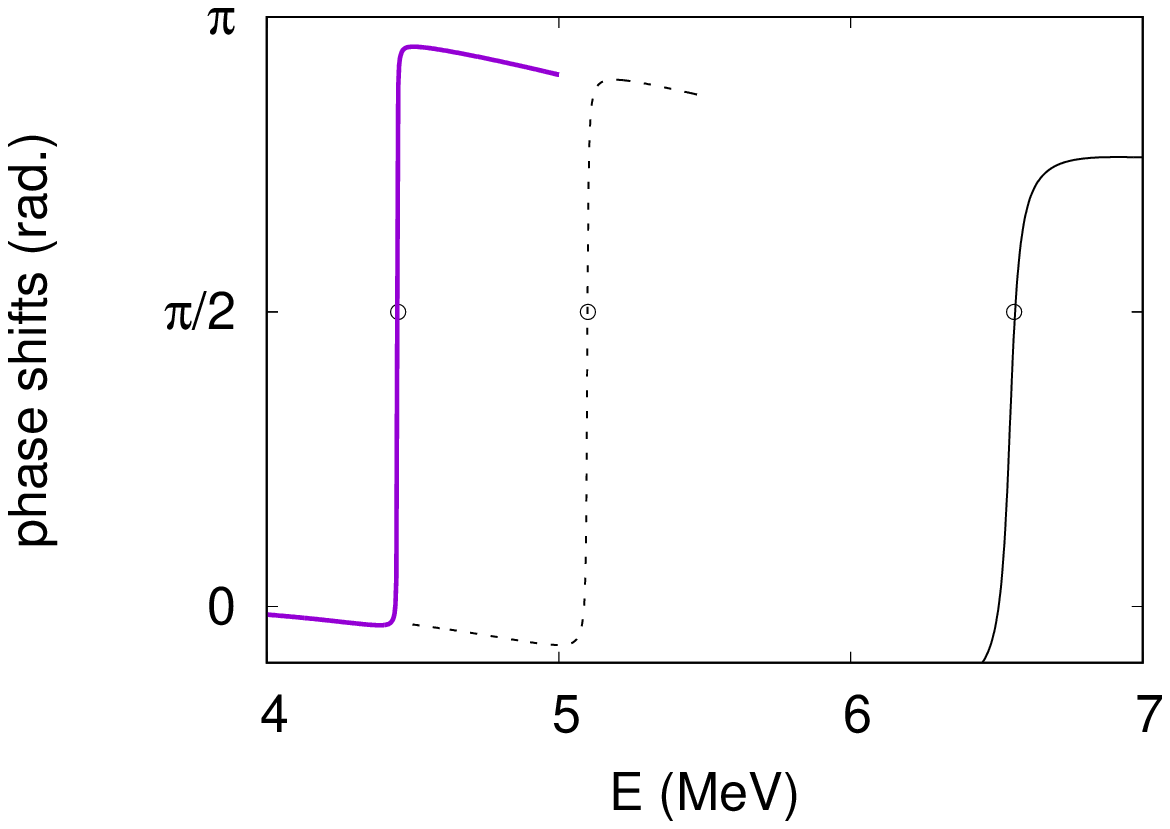}
\caption{(Color online) The scattering phase shifts for the potentials shown in Fig.
\ref{fig-1}. The small circles denote the resonance energies, taken as $\delta(E_R)=\pi/2$,
with values $E_R=4.44$~MeV ($J=0$), $5.09$~MeV ($J=2$) and $6.56$ ~MeV ($J=4$).}
\label{fig-6}       
\end{figure}

These fitted $E_R$ and $\Gamma_R$ values are presented in Table \ref{table1} for grids with
$R_{max} = 3000$ and $7000$ fm. Also shown are the values calculated using a complex-energy
S-matrix pole search and from the scattering phase shifts \cite{Merzbacher}. The phase shift
function, $\delta(E)$, for the three partial waves, is displayed in Fig. \ref{fig-6} and the
associated resonance widths were determined using the formula, $\Gamma_R = 2 \, (d \, \delta
(E)/dE \, |_{E_R})^{-1}$ \cite{Nimrod}, that provides accurate widths for narrow, isolated
resonances \cite{Nimrod}. Evident from Fig. \ref{fig-6} is that, in the present case, the
resonances occur in the presence of significant background phases. Comparing the results of the
present method to those from the S-matrix poles approach, in Table \ref{table1}, we note that
the resonance energy centroids are in excellent agreement (all to better than $1\%$), while
there is a disagreement ($\sim 1$~keV for $J=0$ and $2$) in the widths of the narrower
resonances, using the grid with $R_{max} = 3000$ fm. The latter can be improved by using more
extended grids, as demonstrated for the grid with $R_{max} = 7000$ fm. The mapped Fourier
method for non-uniform grids \cite{MFM} can improve the efficiency of the present approach.
Overall, the present method provides a reliable means to locate and quantify the potential
resonances. The resonances calculated in the present study account for some resonant structures
observed in the astrophysical S-factor for stellar carbon burning \cite{ADTWiescher}.

\begingroup
\begin{table}[ht]
\caption{Resonance energies (MeV) and their widths (keV), extracted from Fig. \ref{fig-5} for
$R_{max} = 3000$ and 7000 fm, are compared to those from the S-matrix pole search and
scattering phase shift methods. While the absolute errors of all $E_R$ from the present
method are $< 0.01\%$, those for $\Gamma_R$ are between $0.5$ and $6\%$.}
\centering
\setlength{\tabcolsep}{10pt}
\renewcommand{\arraystretch}{1}
\begin{tabular}{c c c c c c c}
\hline\hline
& \multicolumn{2}{c} {present method} & \multicolumn{2}{c}{S-matrix poles} & \multicolumn{2}{c}{phase shifts} \\
& \multicolumn{2}{c}{$R_{max}=3000 \,\, (7000)$ fm} & \multicolumn{4}{c}{} \\
$J$ & $E_R$ & $\Gamma_R$ & $E_R$ & $\Gamma_R$ & $E_R$ & $\Gamma_R$\\
\hline
0 & 4.430 (4.431) &  2.75 $\pm$ 0.09 (1.10 $\pm$ 0.07) & 4.437 &  1.55 & 4.437 &  1.58 \\
2 & 5.081 (5.082) &  6.62 $\pm$ 0.15 (5.91 $\pm$ 0.29) & 5.088 &  5.70 & 5.089 &  6.14 \\
4 & 6.535 (6.535) & 49.62 $\pm$ 0.23 (49.72 $\pm$ 0.30) & 6.538 & 49.50 & 6.555 & 73.85 \\ [0.5ex]
\hline
\end{tabular}
\label{table1}
\end{table}
\endgroup

\section{Summary}

In summary, a new method for studying resonance phenomena using wave-packet dynamics has
been presented, based on the powerful window operator method. We demonstrate, in the case of
the astrophysically-important, low-energy $^{12}$C + $^{12}$C system, that the method calculates
resonance energies and widths in agreement with alternative, established methods. The present,
dynamical technique is more generally applicable for the quantitative study of resonance
phenomena in different fields, where particles are temporarily trapped by attractive potential
pockets.

\acknowledgments
The support from the STFC grant (ST/P005314/1) is acknowledged. A.D-T is very thankful to the
ECT* in Trento, where some of the calculations were carried out, for their support during a
research visit in August 2018. We thank Prof. Nimrod Moiseyev for constructive comments.

\end{document}